\documentclass[english,aps,twocolumn]{revtex4}
\usepackage[T1]{fontenc}
\usepackage[latin9]{inputenc}
\usepackage{color}
\usepackage{amsmath}
\usepackage{graphicx}
\usepackage{amssymb}
\usepackage{esint}

\makeatletter
\@ifundefined{textcolor}{}
{%
 \definecolor{BLACK}{gray}{0}
 \definecolor{WHITE}{gray}{1}
 \definecolor{RED}{rgb}{1,0,0}
 \definecolor{GREEN}{rgb}{0,1,0}
 \definecolor{BLUE}{rgb}{0,0,1}
 \definecolor{CYAN}{cmyk}{1,0,0,0}
 \definecolor{MAGENTA}{cmyk}{0,1,0,0}
 \definecolor{YELLOW}{cmyk}{0,0,1,0}
 }

\newcommand{\noshow}[1]{}

\makeatother

\usepackage{babel}

\begin{document}

\preprint{This line only printed with preprint option}

\title{Sensitivity to the initial state of interacting ultracold bosons
in disordered lattices}

\author{Beno\^it Vermersch}

\author{Jean~Claude Garreau}

\affiliation{Laboratoire de Physique des Lasers, Atomes et Mol\'ecules, Universit\'e
Lille 1 Sciences et Technologies, CNRS; F-59655 Villeneuve d'Ascq
Cedex, France}

\homepage{http://www.phlam.univ-lille1.fr/atfr/cq}
\begin{abstract}
We study the dynamics of a nonlinear one-dimensional disordered system
obtained by coupling the Anderson model with the Gross-Pitaevskii
equation. An analytical model provides us with a single quantity globally
characterizing the localization of the system. This quantity obeys
a scaling law with respect to the \emph{width of the initial state},
which can be used to characterize the dynamics independently of the
initial state.
\end{abstract}
\maketitle

\section{Introduction}

The Anderson model, introduced about 50 years ago \cite{Anderson:LocAnderson:PR58},
is the simplest model describing the effects of disorder in a quantum
system in a relatively realistic way. It has however been recognized
rather early that the model probably relied on too strong simplifications
to match the actual behavior of electrons in a crystal: The Anderson
model is a one-particle model at zero temperature. Moreover, the difficulty
of experimental investigations in condensed matter systems \cite{Ramakrishnan:DisorderElectrons:RMP85}
prompted a search for other systems where such effects could be observed
in more favorable conditions, leading to studies of the localization
of electromagnetic waves \cite{Wiersma:LightLoc:N97,Maret:AndersonTransLight:PRL06}
and sound waves \cite{vanTiggelen:AndersonSound:NP08,vanTiggelen:Multifrac:PRL09},
where particle interactions are obviously absent. A breakthrough has
been recently realized by using laser-cooled atoms \cite{Chabe:Anderson:PRL08,Aspect:AndersonBEC:N08,Inguscio:AndersonBEC:N08,Kondov:ThreeDimensionalAnderson:S11,Aspect:Anderson3D:arXiv11,Lignier:Reversibility:PRL05},
a development inscribed in a more general trend of studying many-body
systems with ultracold atoms in optical lattices \cite{Bloch:ManyBodyUltracold:RMP08}.

One of the long-standing questions concerning the Anderson model is
the effect of particle-particle interactions. It was conjectured that
the electron-electron coulomb repulsion would suppress localization.
Including such effects in the model, however, implies going from a
{}``simple'' one-particle picture to a very complex many-body description.
The problem of interactions is very rich in disordered ultracold atomic
systems: For such systems, the interaction strength can be driven
from repulsive to attractive, \emph{via} the so-called Feshbach resonances
\cite{Chin:Feshbach:RMP10}. Moreover, by choosing the atomic species,
the quantum statistics can also be changed from Fermi-Dirac to Bose-Einstein
(see, e.g., \cite{SanchezPalencia:DisorderQGases:NP10} and references
therein). One can conjecture, for example, that attractive interactions
shall \emph{increase} localization in a cold Bose gas, but not in
a Fermi gas. All possible combinations of the sign of interactions
and quantum statistics can be similarly considered~\cite{Crepin:DisorderedOneDimensionalBoseFermi:PRL10}.
Mean-field theories, which simplify considerably the many-body problem,
were generally considered to give a rather poor description of the
electronic many-body problem in a crystal \cite{FetterWalecka:ManyBody:2003}.
However, it became clear in recent years that these theories give,
on the contrary, a very satisfactory description of ultracold Bose
gases with (weak) interactions in a variety of situations \cite{Bloch:ManyBodyUltracold:RMP08}.
This puts into evidence the interest of studying cold atoms in the
presence of disorder.

The first experimental result in this new field was the observation
of {}``dynamical localization'', a manifestation of the Anderson
localization in the momentum space, by Raizen and co-workers, in 1994
\cite{Raizen:LDynFirst:PRL94}. Anderson localization of bosons in
1D \cite{Aspect:AndersonBEC:N08} and 1D localization in the Aubry-Andr\'e
model \cite{Inguscio:AndersonBEC:N08}, have also been observed experimentally,
and, recently, observations of the 3D localization of fermions \cite{Kondov:ThreeDimensionalAnderson:S11}
and bosons \cite{Aspect:Anderson3D:arXiv11} were claimed. The effect
of interactions is also being studied experimentally \cite{Inguscio:AubryAndreInteractions:NP10,Inguscio:InteracBosons:NJP11,Inguscio:SubdiffusionInteract:PRL11}.
Finally, the Anderson metal-insulator transition is being actively
studied with a cold atom {}``quantum simulator'' of the 3D Anderson
model \cite{Casati:IncommFreqsQKR:PRL89}, including, the experimental
determination of its critical exponent \cite{Chabe:Anderson:PRL08,Lemarie:AndersonLong:PRA09,Lopez:ExperimentalTestUniversality:arXiv11}
and the study of its critical state \cite{Lemarie:CriticalStateAndersonTransition:PRL10}.

\section{Frame and scope\label{sec:Frame-and-scope}}

We start by considering the tight-binding description of an one-dimensional
(ordered) lattice of period $d$ obtained by projecting the eigenstates
of the spatially periodic Hamiltonian on a basis of localized functions,
usually Wannier functions $w_{n}(x)$, associated with each site $n$:
\begin{equation}
\psi_{\varepsilon}(x)=\sum_{n}c_{n}w_{n}(x),\label{eq:WannierDecomposition}\end{equation}
which produces a discretized eigenvalue problem\[
V_{n}c_{n}+\sum_{r\neq0}T_{r}c_{n+r}=\varepsilon c_{n}.\]
We use here the (usual) symmetric first-neighbors approximation $T_{r}=-T\delta_{r,\pm1}$
and, as usual, rescale time such that $t\rightarrow\hbar t/T$ (and
energies correspondingly: $E=\varepsilon/T$) in order to obtain the
tight-binding equation:\begin{equation}
v_{n}c_{n}-c_{n-1}-c_{n+1}=Ec_{n}.\label{eq:Htb}\end{equation}
If all sites have the same diagonal energy, one can redefine the energy
origin so that $v_{n}\equiv0$, then \cite{Luck:SystDesord:92,AshkroftMermin:SolidStatePhys:76}:
\begin{eqnarray}
E(q) & = & -2\cos q\label{eq:BandTB}\\
\rho(q) & = & 1/\pi\label{eq:DensityTB}\end{eqnarray}
where $q$ is the quasimomentum, $E(q)$ are the energies on the first
band, $\rho$ is the density of states. The eigenfunctions $\psi_{q}(x)$
are delocalized Bloch waves of quasimomentum $q\in\left[0,\pi\right]$.

The Anderson model postulates that the main effect of crystalline
disorder is to randomize the $v_{n}$ in eq.~(\ref{eq:Htb}) in an
interval $\left[-W/2,W/2\right]$, an approximation known as {}``diagonal
disorder''. The introduction of disorder\textcolor{green}{{} }redistributes
the eigenenergies between the values $-2-W/2$ and $2+W/2$ and, more
importantly, produces a \emph{localization} of the corresponding eigenfunctions
\footnote{In one dimension, \emph{all} eigenfunctions are localized, whatever
the disorder $W$.%
}. We shall index the eigenfunctions by $\nu=q-\pi/2$ so that $\nu=0$
is at the center $q=\pi/2$ of the energy band. The shape of a given
eigenfunction depends on $E_{\nu}$ and on the realization $\{v_{n}\}$
of the disorder, but if one averages the eigenfunctions in a given
energy interval $[E,E+\Delta E]$ over many realizations of the disorder
one finds an exponential shape\[
\overline{\psi_{\nu}}(n)\sim\exp\left(-\frac{\left|n-n_{\nu}\right|}{\ell_{\nu}(W)}\right)\]
(we use overbars to indicate averages over realizations of the disorder)
where $\ell_{\nu}(W)$ is the \emph{localization length}. In the 1D
case and in the limit of weak disorder one can show that \cite{Luck:SystDesord:92}\begin{equation}
\ell_{\nu}(W)\approx\frac{96}{W^{2}}\cos^{2}\nu,\label{eq:LocalizationLenAnderson}\end{equation}
which vanishes at the borders of the band and has its maximum value
$\ell_{0}\sim96W^{-2}$ at the center of the band. This localization
can be interpreted by noting that, in the absence of disorder, all
sites have identical energies, so the particle can tunnel from a site
to the next one, generating a diffusive motion which leads, asymptotically,
to a complete delocalization in the lattice. In the presence of diagonal
disorder, however, it is unlikely that neighbor sites have close enough
energies, so that in general, the particle can only perform \emph{virtual}
transition to neighbor sites, which leads to a localized exponential
spreading of the eigenfunction.

We take into account interactions by using a mean-field approximation.
The decomposition eq.~(\ref{eq:WannierDecomposition}) is used to
transform %
\footnote{In order to do so one generalizes the first-neighbors approximation
by supposing that $\int dxw_{n}^{3}(x)w_{n\pm1}(x)\ll\int dxw_{n}^{4}(x)$
and keeps only the {}``diagonal'' nonlinear term.%
} the Gross-Pitaevskii equation (or nonlinear Schr\"odinger equation)
in a discrete nonlinear set of equations\begin{equation}
i\dot{c}_{n}=v_{n}c_{n}-c_{n-1}-c_{n+1}+g\left|c_{n}\right|^{2}c_{n}\label{eq:DANSE}\end{equation}
where $g$ is the parameter characterizing atom-atom interactions,
proportional to the $s$-wave diffusion length \cite{Stringari:BECRevTh:RMP99,PethickSmith:BoseEinstein:08}.
Disorder is introduced by randomizing $v_{n}$, as before. This model
is occasionally named DANSE (Discrete Anderson Nonlinear Schr\"odinger
Equation). As we use mean-field theory throughout this work, we shall
use the terms {}``interaction'' and {}``nonlinearity'' interchangeably.

The main question in the problem of interacting disordered systems
is, will interactions destroy localization? The DANSE model has been
used, e.g. by Pikovskii and Shepelyansky \cite{Shepelyansky:DisorderNonlin:PRL08},
to investigate this question. They observed numerically a revival
of diffusion, leading, at very long times ($\sim10^{7}$), to a $\left\langle x^{2}\right\rangle \sim t^{\alpha}$
sub-diffusive behavior with $\alpha\approx0.3$. Other works also
tackled this problem, with somewhat contradictory results. For example,
in~\cite{FlachAubry:DisorderNonlin:PRL08}, the use of the so-called
{}``participation number'' $\left(\sum_{n}\left|c_{n}\right|^{4}\right)^{-1}$
(instead of $\left\langle x^{2}\right\rangle $) as the quantity characterizing
diffusion, lead to the conclusion that localization survives even
in presence of a moderate nonlinearity, but subsequent studies \cite{Flach:DisorderNonlin:PRL09,Flach:DisorderNonlineChaos:EPL10}
confirmed the existence of a subdiffusive dynamics, with an exponent
$\alpha$ that typically depends on the initial state. Experimental
evidence of subdiffusion in the Aubry-Andr\'e model has also been observed
recently \cite{Inguscio:SubdiffusionInteract:PRL11}.

One can formally write the ``diagona''l part of eq.~(\ref{eq:DANSE})
as $\left(v_{n}+g\left|c_{n}\right|^{2}\right)c_{n}$, which allows
us to interpret the nonlinear term $v_{n}^{NL}\equiv g\left|c_{n}\right|^{2}$
as a {}``correction'' (depending on the site population) to the
energy of site $n$ %
\footnote{Note that $\left\langle v_{n}^{NL}\right\rangle \approx\mu,$ where
$\mu$ is the chemical potential.%
}. This heuristic picture, although not rigorous, is often useful in
interpreting the behavior of the system.

\begin{figure*}[t]
\begin{centering}
\includegraphics[width=0.65\columnwidth]{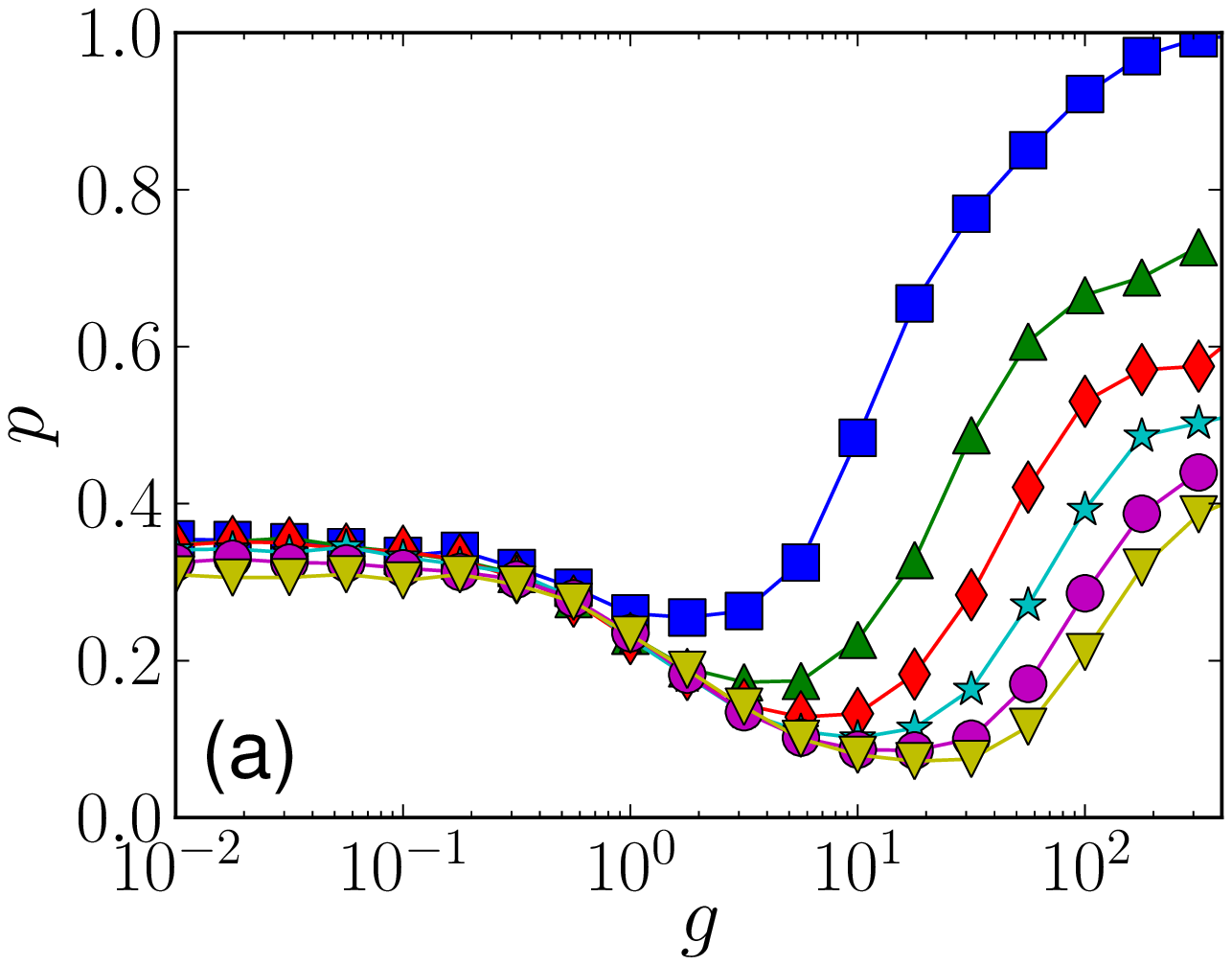}\quad{}\includegraphics[width=0.65\columnwidth]{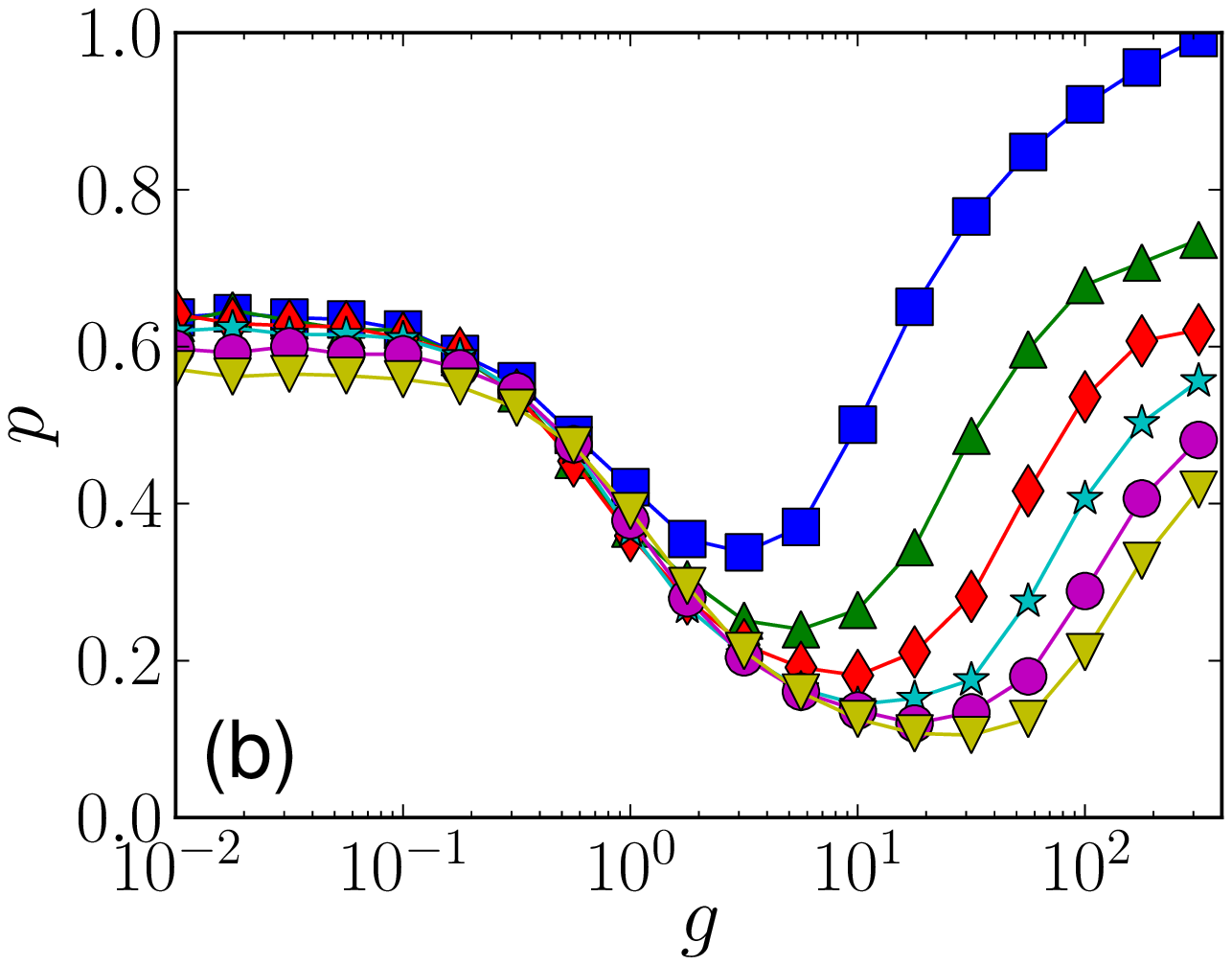}\quad{}\includegraphics[width=0.65\columnwidth]{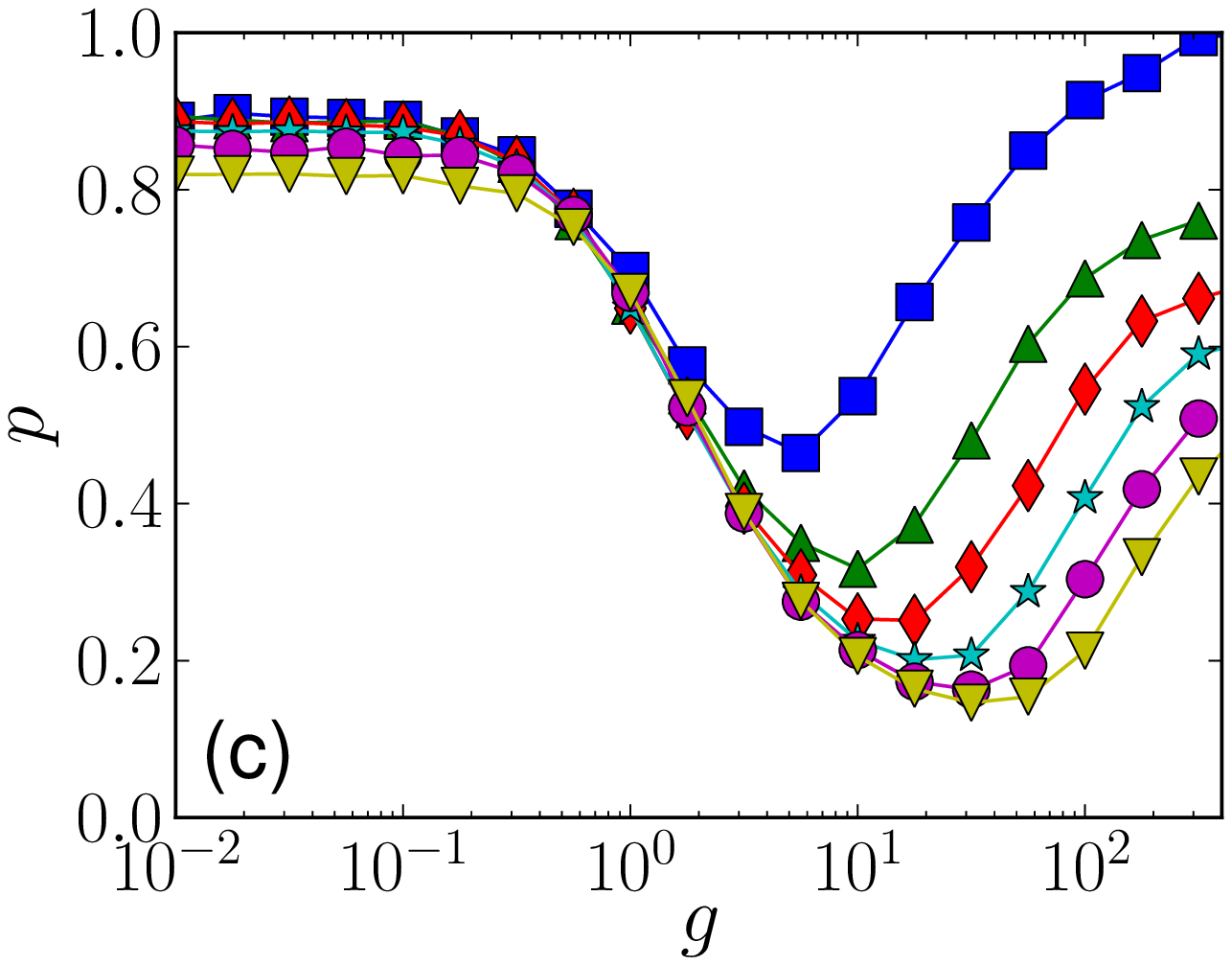}
\par\end{centering}

\caption{\label{fig:Lossesvsg}Survival probability $p(g,t=10^{5})$ as a function
of the nonlinear parameter $g$ for different values of the width
$L_{0}$ of the initial state and disorder (a) $W=2$, (b) $W=3$,
(c) $W=4$. Three regimes can be identified: Quasi-localized for low
$g$; chaotic evolution with destruction of the localization for intermediate
values of $g$; and self-trapping for high $g$. The nonlinearity
destroys the localization almost completely in the chaotic regime,
and self-trapping is more efficient, as expected, for smaller values
of the initial state width $L_{0}$. For low disorder and small initial
state width the localization due to self-trapping becomes much more
efficient than the Anderson localization. Results are averaged typically
over 1000 realizations of the disorder and of the initial phase distribution.
Values of $L_{0}$: 3 (blue squares), 7 (green triangles), 13 (red
diamonds), 21 (cyan stars), 31 (magenta circles), 41 (yellow inverted
triangles).}
\end{figure*}

\begin{figure}
\begin{centering}
\includegraphics[width=0.7\columnwidth]{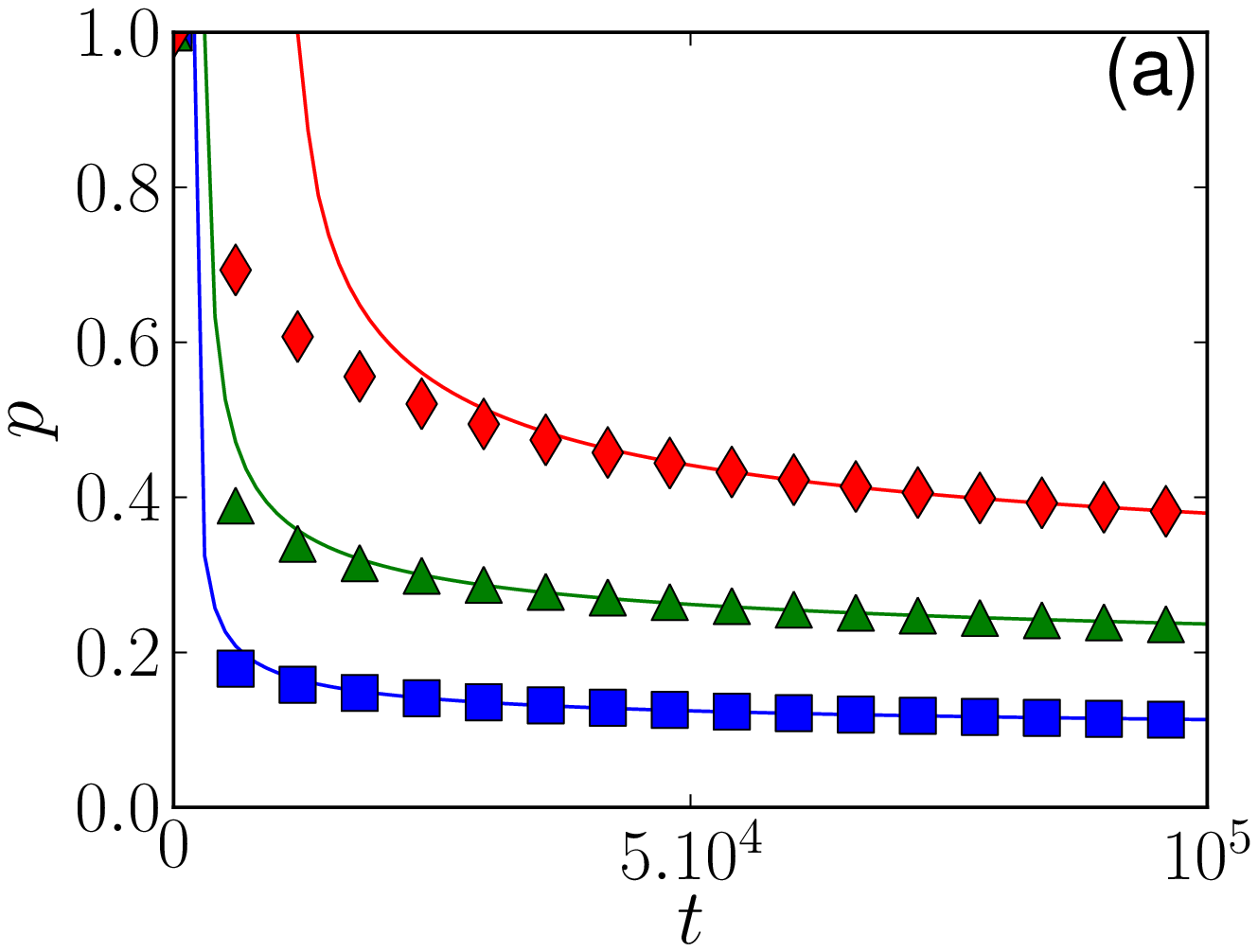}\\
\includegraphics[width=0.7\columnwidth]{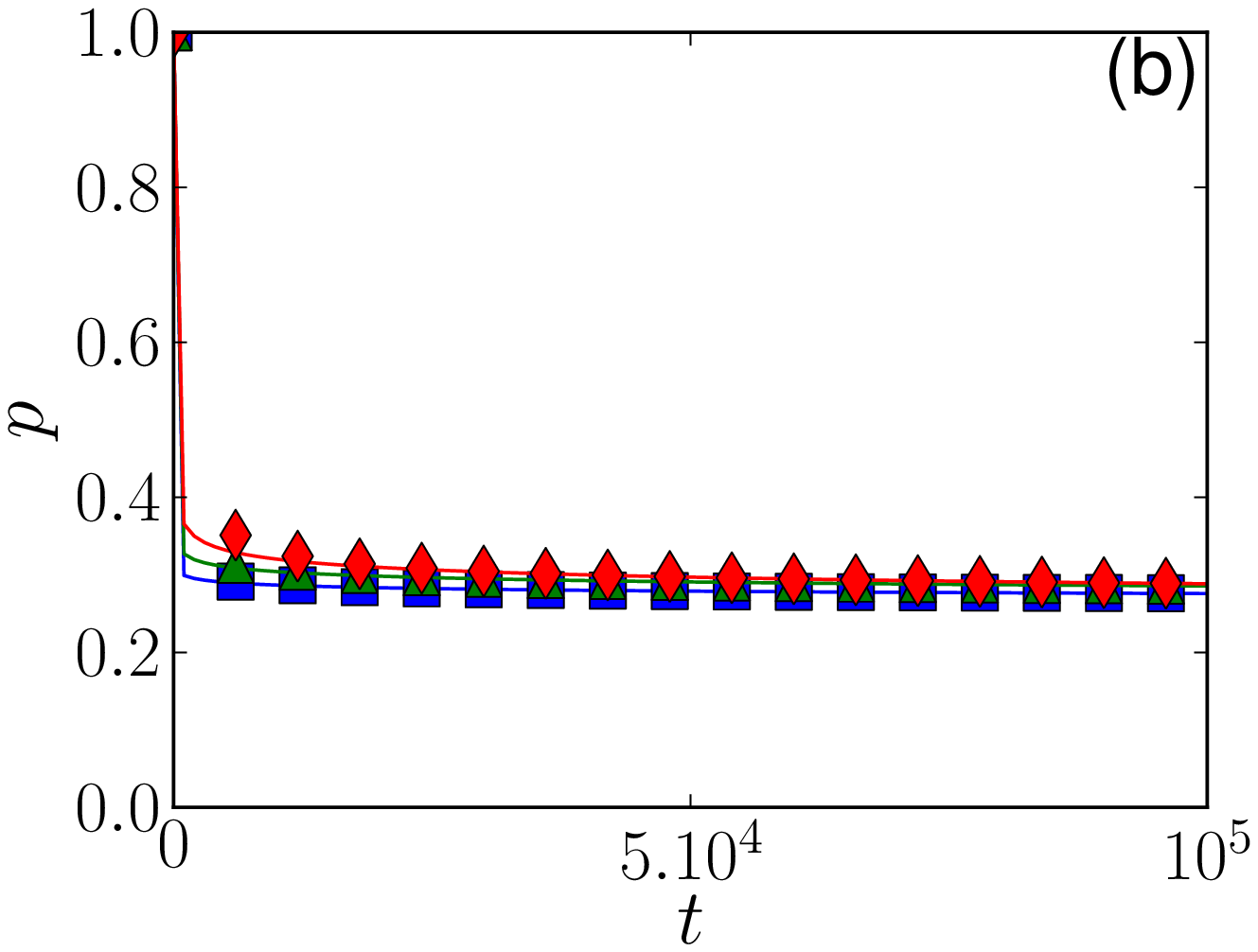}
\par\end{centering}

\caption{\label{fig:pvst_gneq0}Comparison between numerical simulations (symbols)
of $p(g,t)$ and the fit (solid lines) by eq.~(\ref{eq:p(g,t)})
for $L_{0}=31$ and (a) $g=1$, (b) $g=100$. Values of $W$: 1 (blue
squares), 2 (green triangles) and 3 (red diamonds).}
\end{figure}

Mathematically, much is known about eq.~(\ref{eq:DANSE}) in the
absence of disorder, in different contexts~(see e.g.~\cite{SulemSulem:NonlinSchroedingerEq:99}):
As a special case of the Ginsburg-Landau equation \cite{Annett:SupercondSuperfluidConds:04}
it can describe some aspects of superfluidity, and, in optics, it
describes the Kerr effect in a multimode system~\cite{Yariv:OptElectron:91}.
Two nonlinear effects play a particularly important role in our study.
The first one is the called \emph{self-trapping}, which manifests
itself when a given site $n$ has a much larger population than its
neighbors. In such case, the correction $v_{n}^{NL}$ decouples it
from its neighbors (much as the disorder itself does), thus inhibiting
diffusion. The second effect is the existence of chaotic behaviors
(in a classical sense -- that is, chaos related to sensitivity to
initial conditions), due to the presence of nonlinearity \cite{Thommen:ChaosBEC:PRL03,Smerzi:InstabilityBEC:PRL04,Inguscio:InstabilityBEC:PRL04}.
The chaotic evolution of the amplitudes $c_{n}$ may generate strong
variations of the nonlinear correction $v_{n}^{NL}$, eventually bringing,
even if only for a short time, neighbor sites close to degeneracy,
thus favoring diffusion. In the presence of both disorder and interactions,
different regimes are possible: Localization is expected to survive
(or at least be destroyed only at very long times) if interactions
are weak enough compared to disorder, diffusion (or subdiffusion)
induced by the chaotic evolution is expected if these two effects
are of the same order of magnitude, and self-trapping is expected
to dominate, inhibiting diffusion, in the regime of strong interactions.
These regimes have indeed been observed numerically~\cite{Flach:DisorderNonlineChaos:EPL10,Flach:DisorderNonlin:PRL09}.
It can be tempting to classify these regimes by comparing $v_{n}^{NL}$
to $W$, but one must not forget that, as $v_{n}^{NL}$ depends on
$c_{n}(t)$, the dynamics \emph{depends on the particular trajectory
of the system in the $c_{n}$ space}. Such a classification scheme
is thus useless to define {}``phases'' of the system. 

This is an important and - from the point of view of the familiar
\emph{linear} quantum systems -- unusual characteristic of nonlinear
systems: The initial state plays a very important role (much more
than in linear systems) in determining the dynamics. For example,
if one takes, as in \cite{Shepelyansky:DisorderNonlin:PRL08}, an
initial condition $c_{n}(t=0)=\delta_{n0}$, one strongly favors self-trapping,
and thus reduced diffusion. A choice like $c_{n}=1/\sqrt{N}$ for
$|n|<N/2$ favors, for moderate values of $N$, chaotic dynamics,
and thus diffusion. The main goal of the present work is to study
the impact of initial conditions in the dynamics of the DANSE model.
We shall show that this generally forbids a general classification
of the dynamics in the parameter plane $g,W$. In the frame of a simplified
model, introduced in sec.~\ref{sec:The-model}, we shall however
show that one can define an {}``effective wavepacket length'' obeying
scaling laws \emph{with respect to the extent of the initial state}
(sec.~\ref{sec:ScalingLaws}). This allows us to define a generalized
nonlinearity parameter, depending on the initial state, which leads
to a more satisfactory classification of the dynamical behavior in
the weak disorder limit. Finally, we discuss in sec.~\ref{sec:OtherEffects}
the impact of some important neglected effects on our results.

\begin{figure*}
\begin{centering}
\includegraphics[width=0.65\columnwidth]{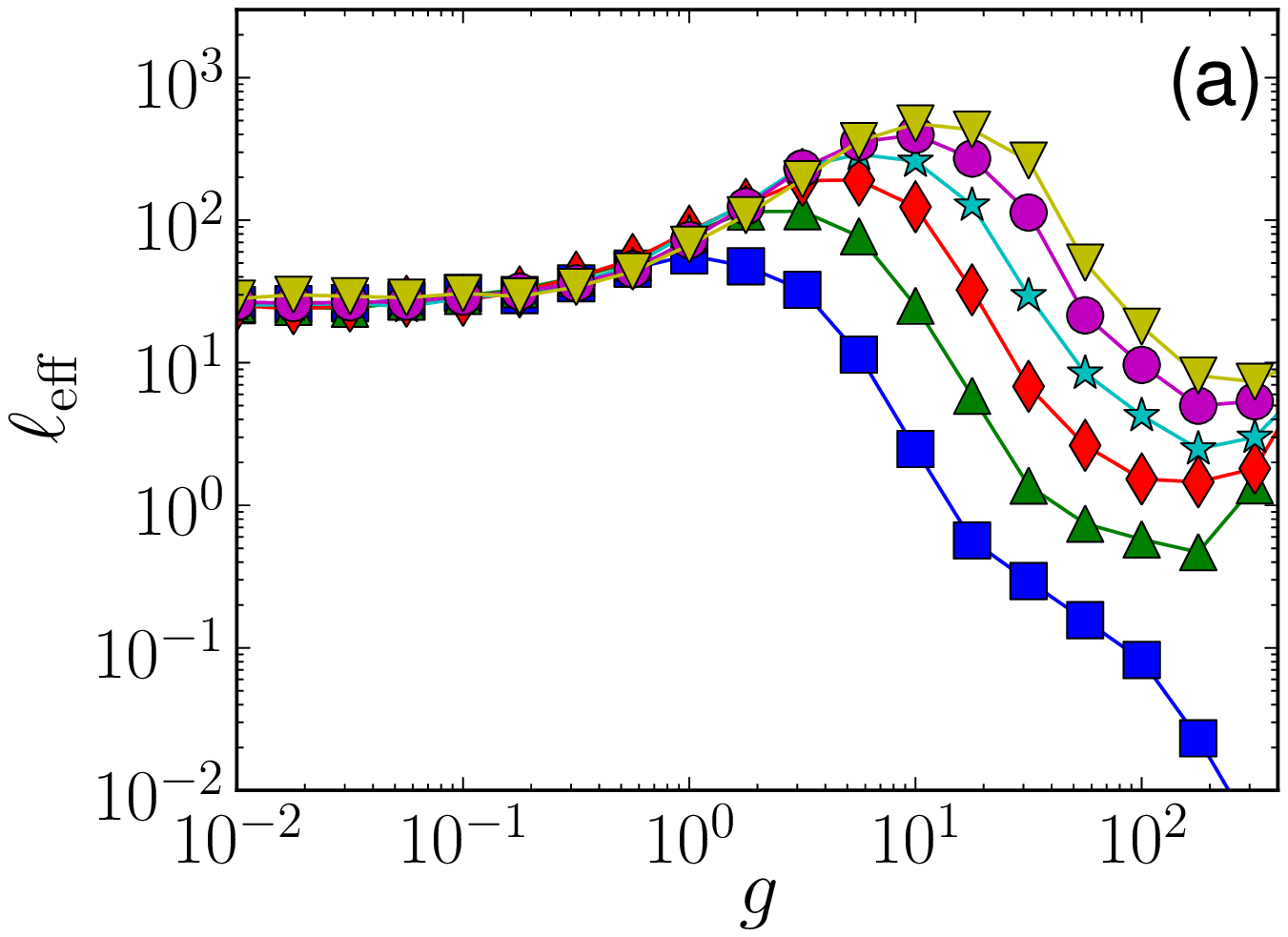}\quad{}\includegraphics[width=0.65\columnwidth]{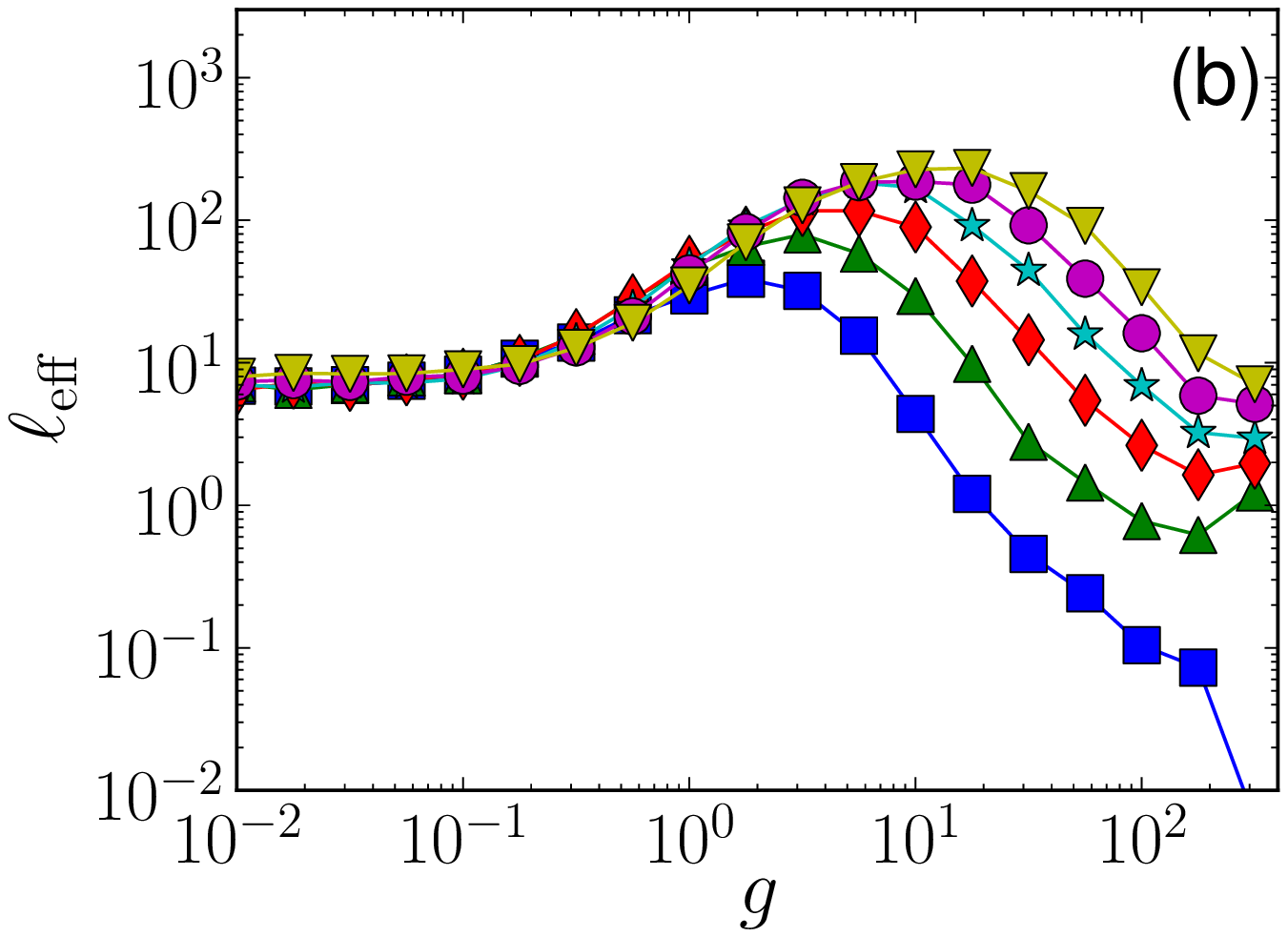}\quad{}\includegraphics[width=0.65\columnwidth]{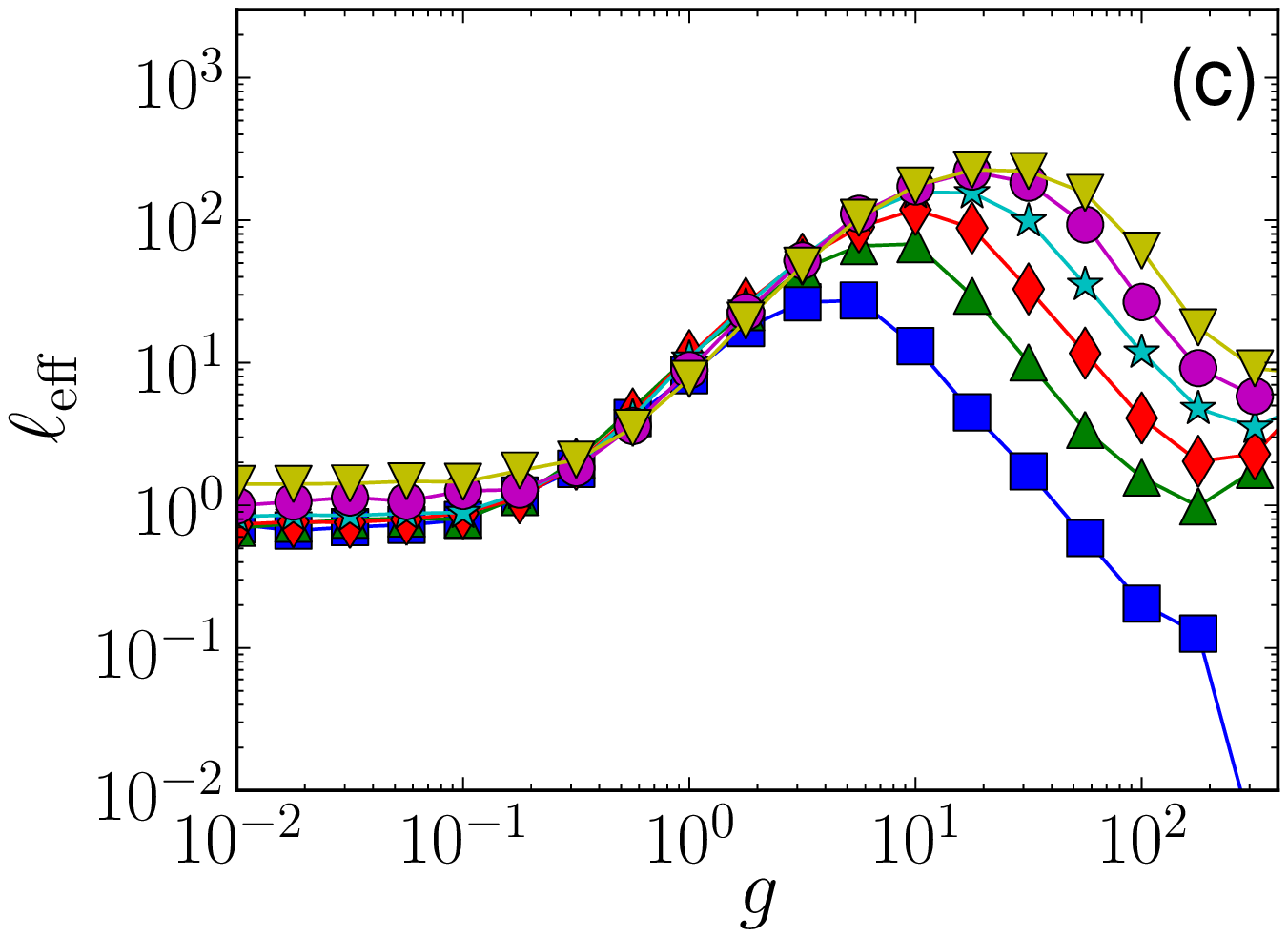}
\par\end{centering}

\caption{\label{fig:leffvsg}Effective localization length as a function of
the nonlinearity $g$, for (a) $W=2$, (b) $W=3$ and (c) $W=4$ and
various values of the width $L_{0}$ of the initial state. Note that
for $g\rightarrow0$ we retrieve the linear localization length $\ell_{0}\approx96W^{-2}$
in the weak disorder limit $W\le3$. Same graphic conventions as in
fig.~\ref{fig:Lossesvsg}.}
\end{figure*}

\begin{figure}
\begin{centering}
\includegraphics[width=0.7\columnwidth]{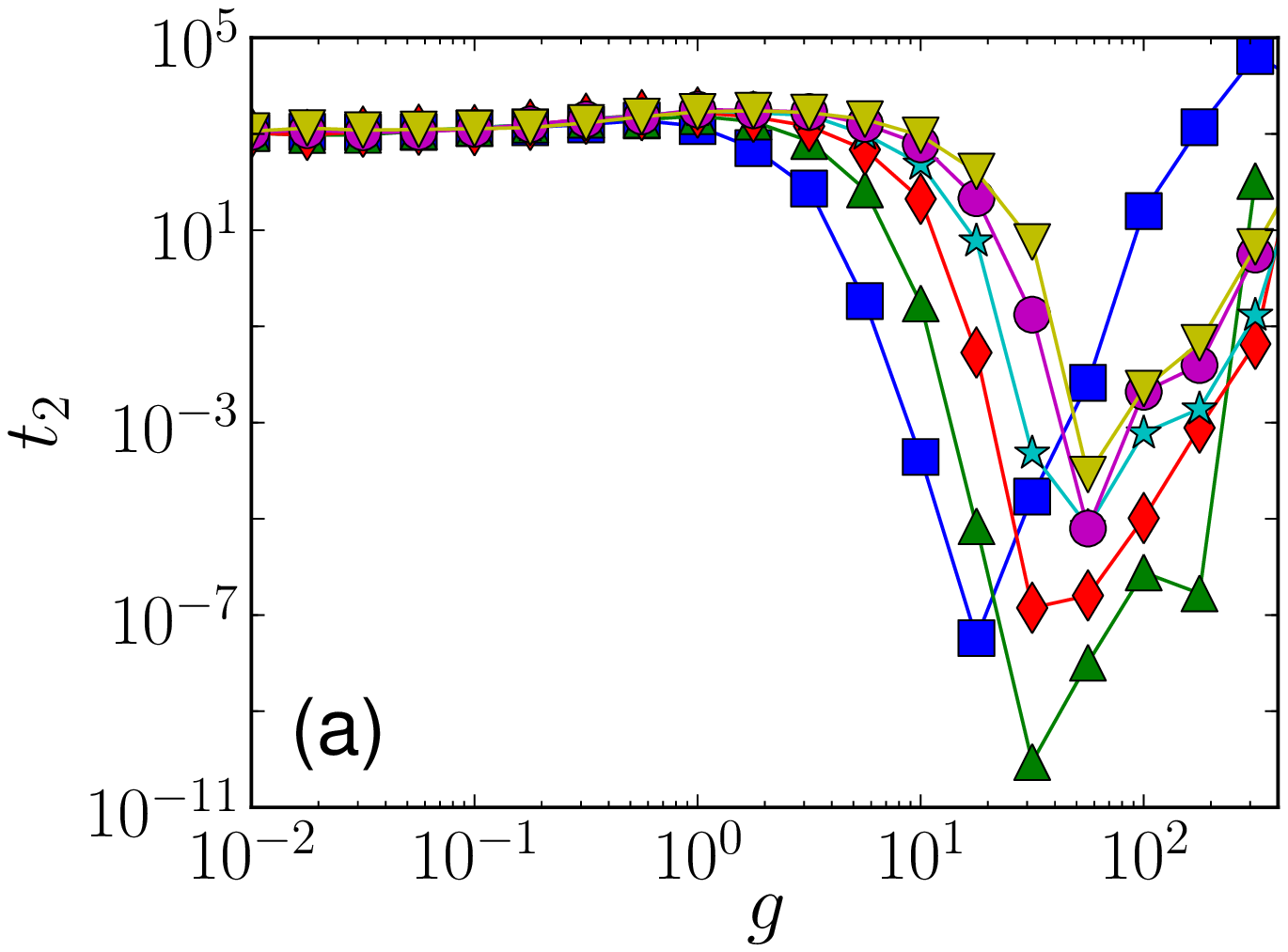}\\
\includegraphics[width=0.7\columnwidth]{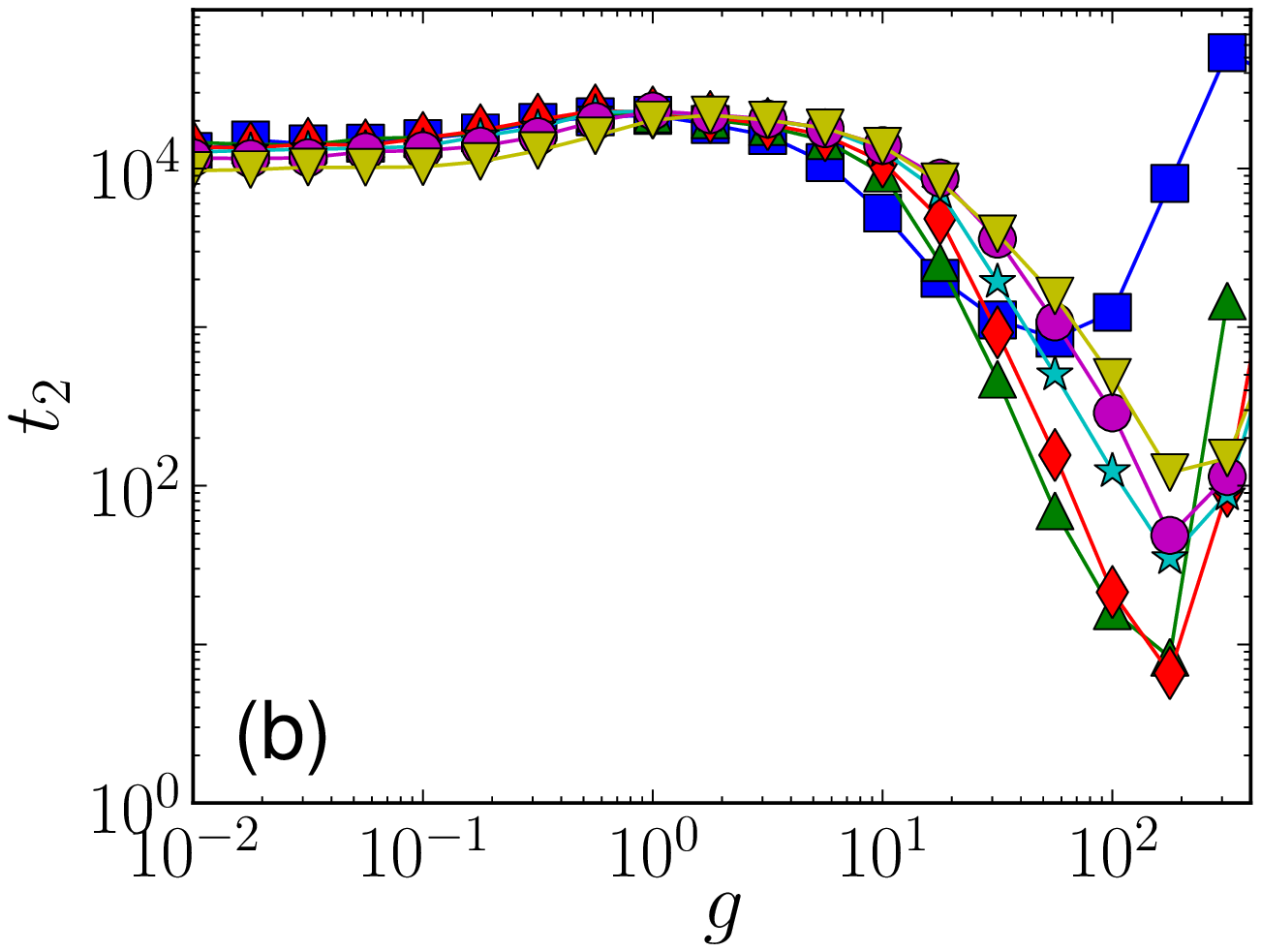}
\par\end{centering}

\caption{\label{fig:t2vsg}Activation time $t_{2}$ as a function of the nonlinearity
$g$ for (a) $W=2$ and (b) $W=4$ and various values of the width
$L_{0}$ of the initial state. Same graphic conventions as fig.~\ref{fig:Lossesvsg}.}
\end{figure}

\section{The model\label{sec:The-model}}

The search for asymptotic behaviors of the type $\left\langle x^{2}\right\rangle \sim t^{\alpha}$
in the DANSE model implies describing the system by an array whose
size increases with the evolution, the number of elements typically
increasing as $t^{\alpha/2}$. This is not our aim here: If one starts
with an initial state localized in a relatively small region, the
restoration of (sub-)diffusive dynamics by the nonlinearity should
be apparent on the fact that parts of the wavepacket must continuously
{}``escape'' the initial region. We thus study a disordered lattice
in a box $\mathrm{L}$ containing $L\sim100$ sites, starting from
an initial state of width $L_{0}$. At the frontiers $x=\pm L/2$
of this box we place {}``absorbers'' by adding an imaginary part
$-i\eta_{a}$ to the potential, which increases exponentially over
a distance $L_{a}\sim10$, in order to prevent reflection of the parts
of the wavepacket approaching the limits of the box. 

We are interested here in the effect of the spatial extension of the
initial wavepacket. In order to simplify the problem, we use a initial
wavepacket of square shape and width $L_{0}$:\begin{equation}
\left|c_{n}(0)\right|=\begin{cases}
\left(L_{0}\right)^{-1/2} & |n|\le\left(L_{0}-1\right)/2\\
0 & \mathrm{otherwise.}\end{cases}\label{eq:InitialState}\end{equation}
We prevent dominant quantum interference effects by setting random
phases to the $c_{n}$. We shall consider in sec.~\ref{sec:OtherEffects}
the effect of non-random quantum phases. 

\begin{figure*}[t]
\begin{centering}
\includegraphics[width=0.65\columnwidth]{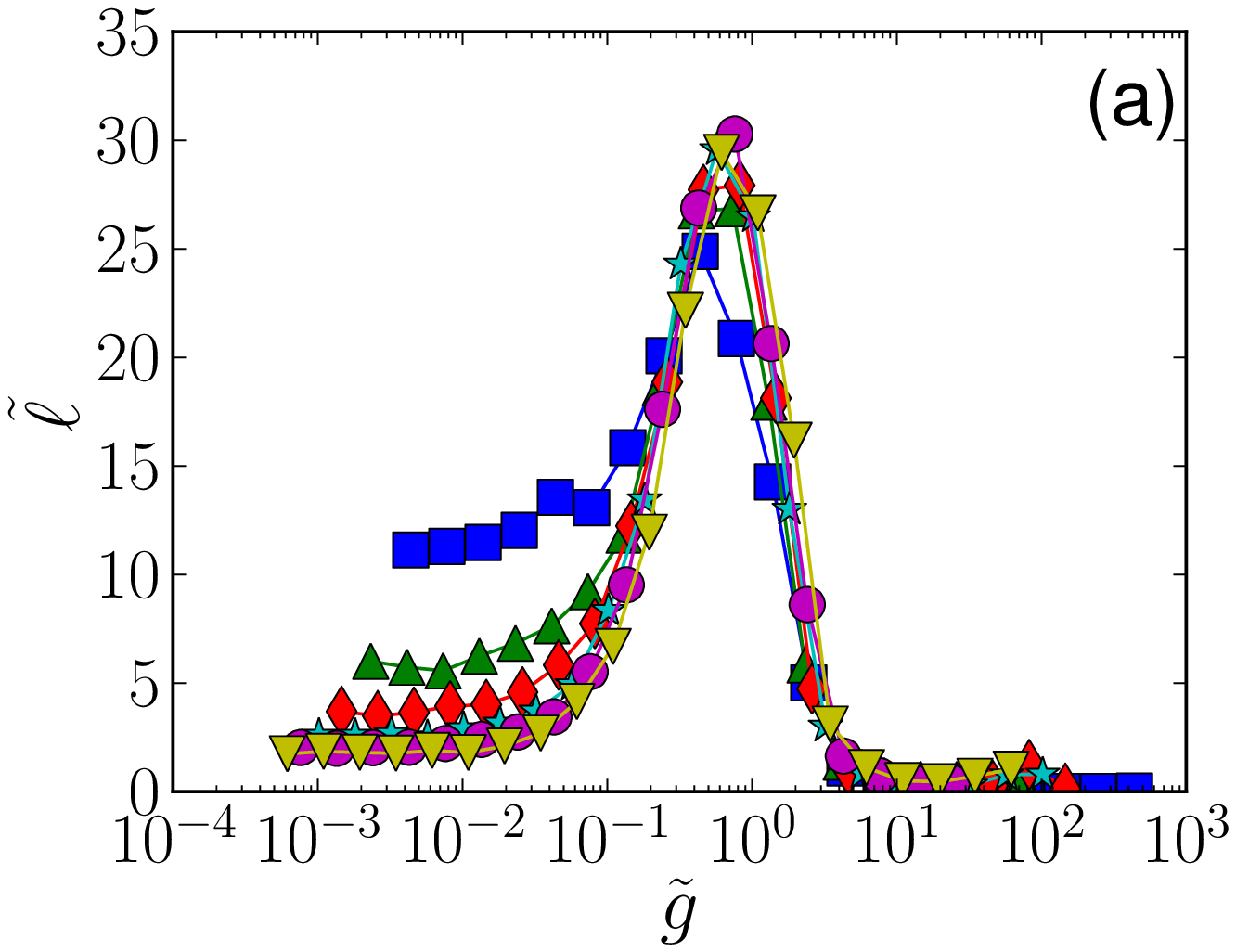}\quad{}\includegraphics[width=0.65\columnwidth]{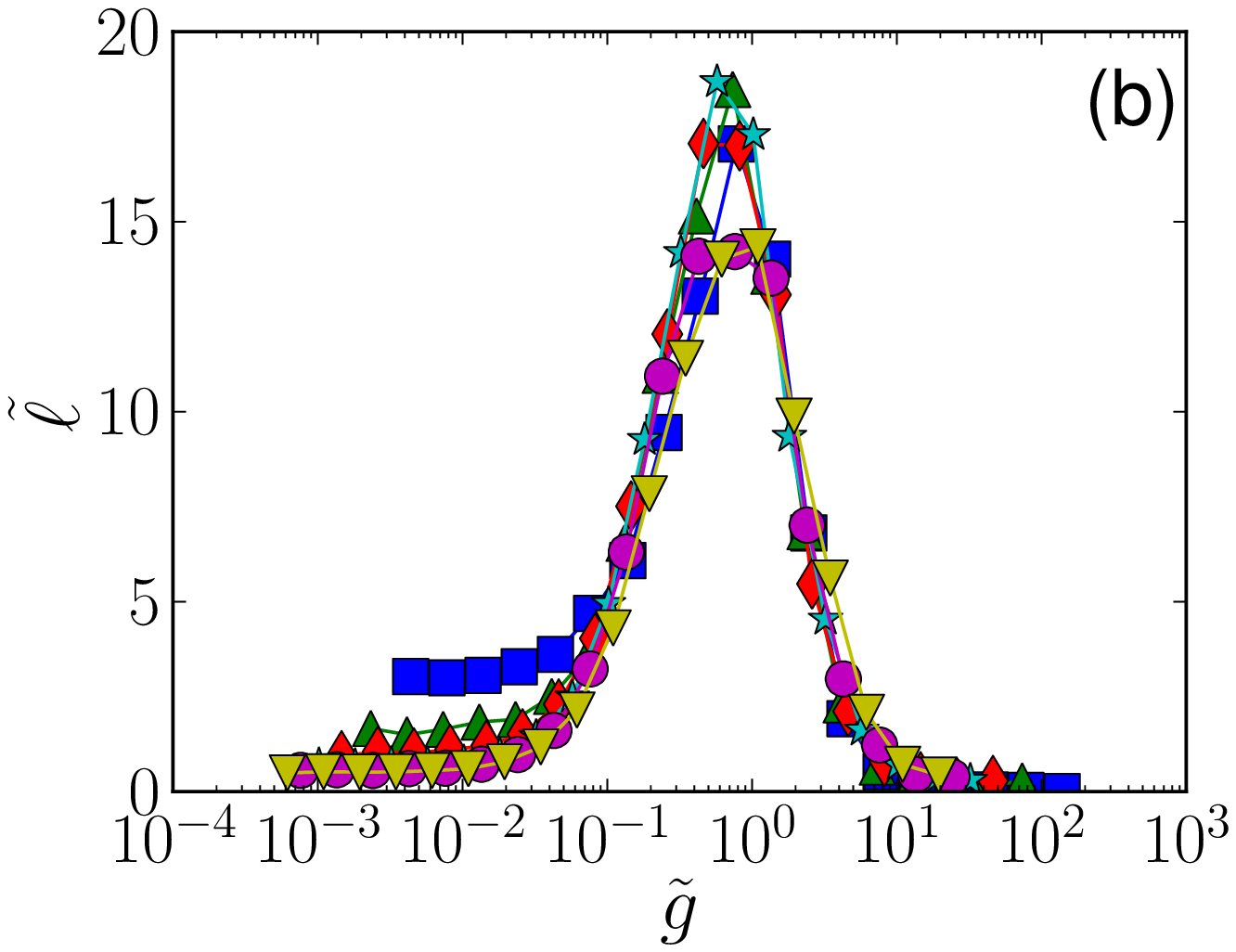}\quad{}\includegraphics[width=0.65\columnwidth]{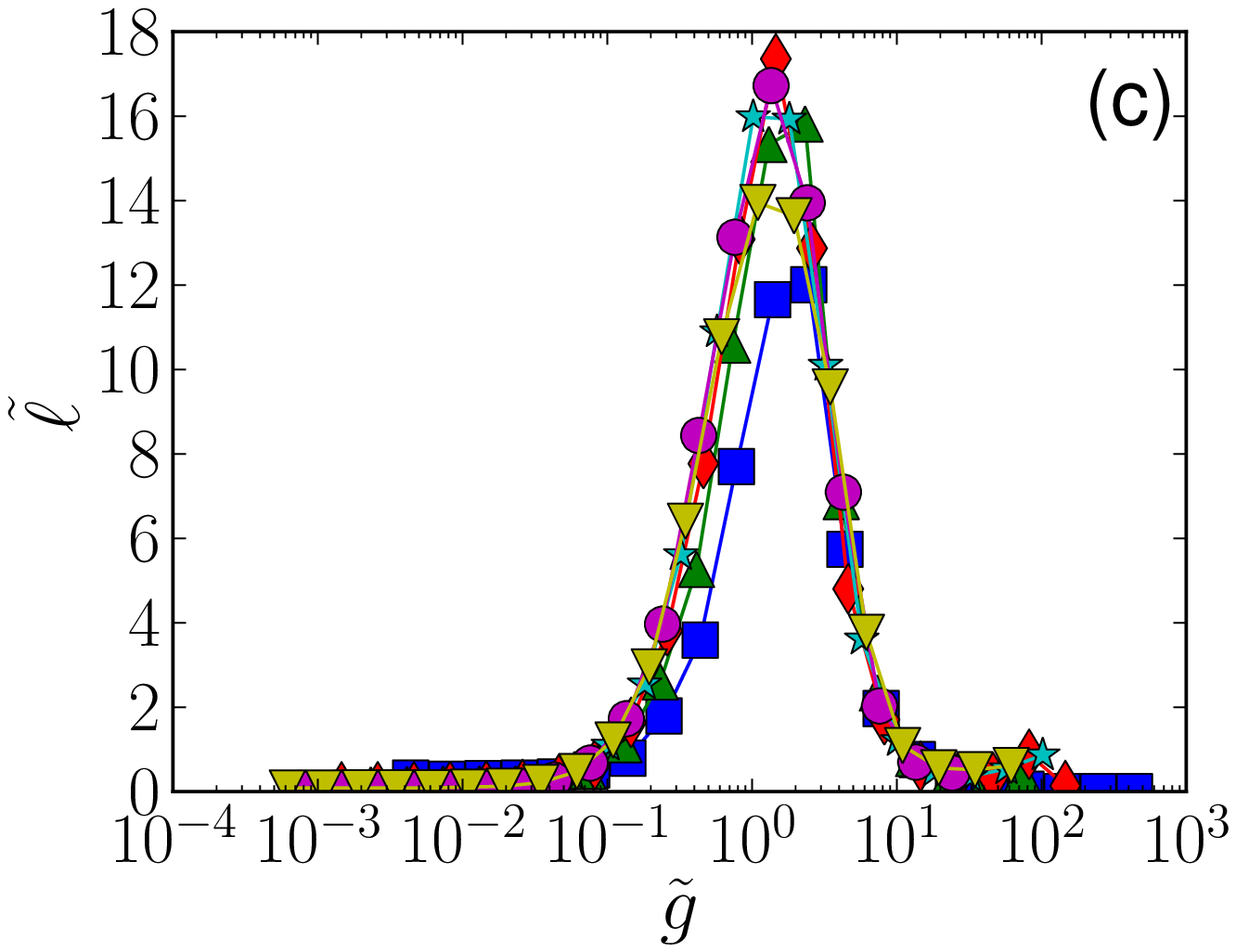}
\par\end{centering}

\caption{\label{fig:Scalinglaw}Scaling law. The same data of fig.~\ref{fig:leffvsg}
is plotted in terms of the scaled quantities $\widetilde{g}=gL_{0}^{-3/4}$
and $\widetilde{\ell}=\ell_{\mathrm{eff}}L_{0}^{-3/4}$ for (a) $W=2$,
(b) $W=3$ and (c) $W=4$. A clear grouping of the curves in the nonlinear
region is observed. Same graphic conventions as fig.~\ref{fig:Lossesvsg}.}
\end{figure*}

\begin{figure*}
\begin{centering}
\includegraphics[clip,width=0.95\columnwidth]{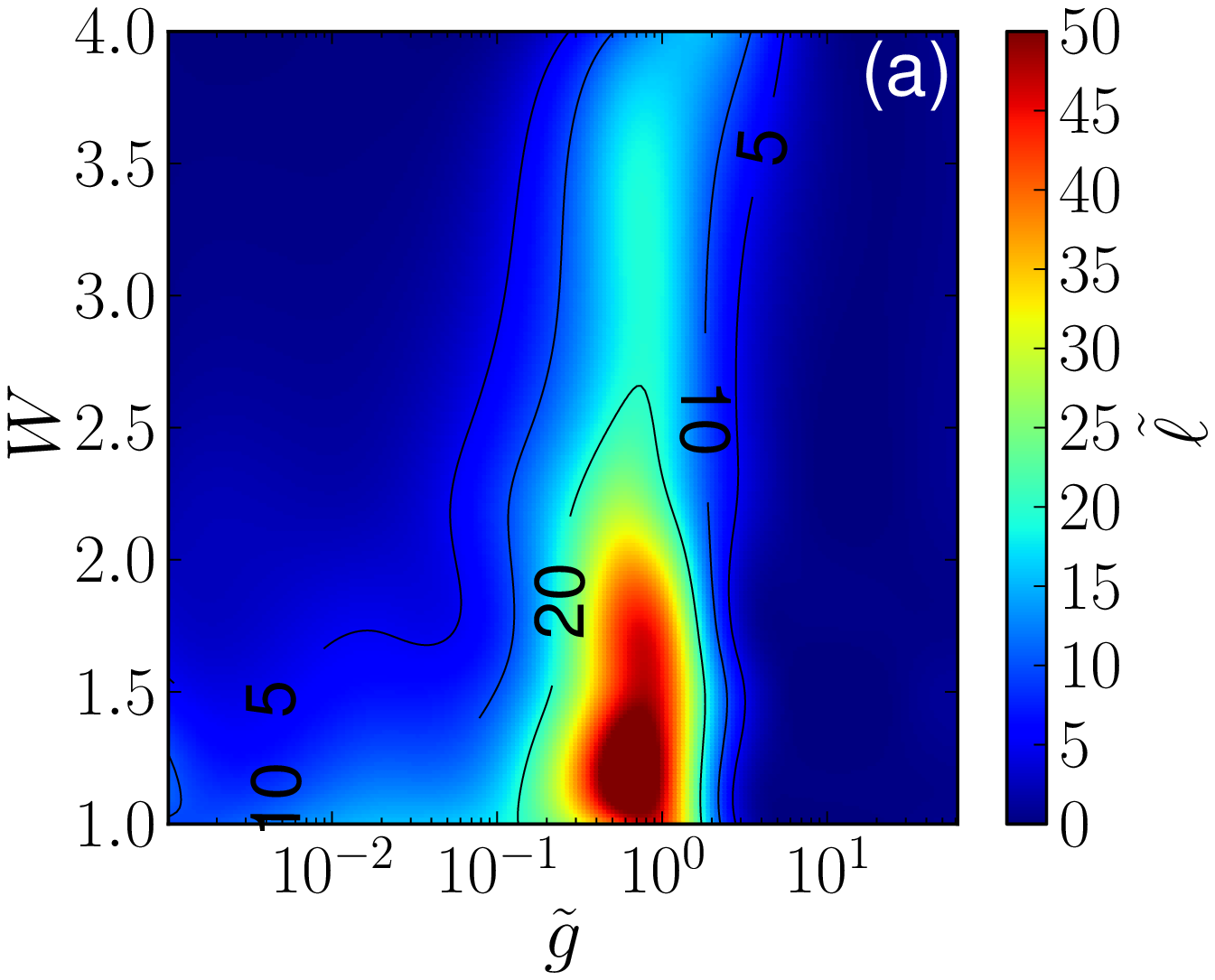}\includegraphics[clip,width=0.95\columnwidth]{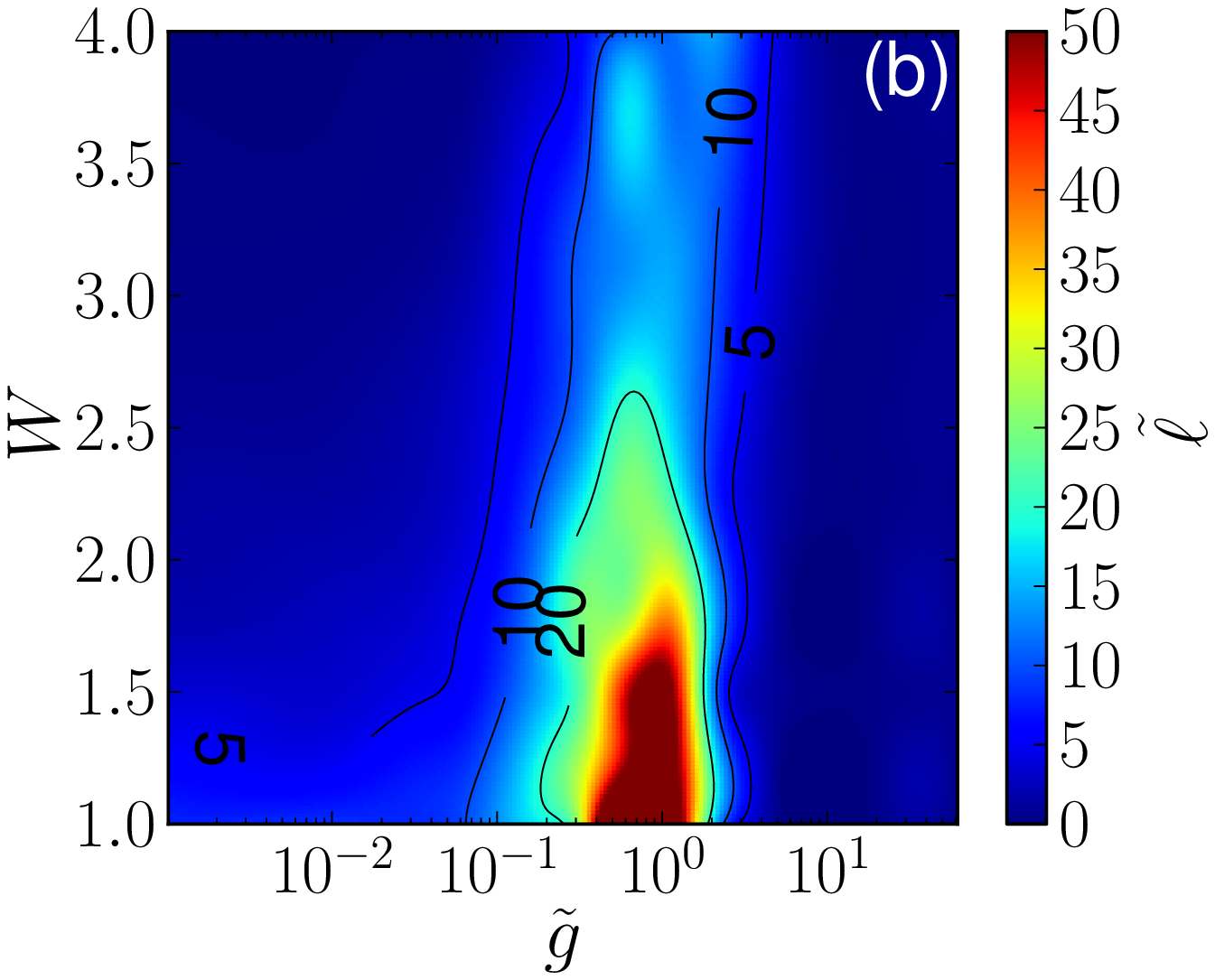}
\par\end{centering}

\caption{\label{fig:leffvsgtildeW} False colors plot of the scaled effective
wavepacket length $\tilde{\ell}$ as a function of the scaled nonlinearity
$\tilde{g}$ and of the disorder $W$ for (a) $L_{0}=21$ and (b)
$L_{0}=41$. The structure of the parameter space is very similar
for the two values of the initial state width.}
\end{figure*}

The part of the wavepacket that, at time $t$, has remained in the
box $\mathrm{L}$ (that we shall call the {}``survival probability'')
is given by\[
p(g,t)=\sum_{n=-(L-1)/2}^{(L-1)/2}\left|c_{n}(t)\right|^{2}.\]
We display in fig.~\ref{fig:Lossesvsg} $p(g,t)$ calculated at $t=10^{5}$
as a function of the nonlinearity $g$ for various sizes of the initial
state $L_{0}$, for a disorder amplitude $W=2$ (part (a) of fig.~\ref{fig:Lossesvsg}),
$W=3$ (b) and $W=4$ (c). The numerical integration is done by the
Crank-Nicholson method~\cite{NumRec:07}.\textcolor{blue}{{} }The probability
density $|c_{n}|^{2}$ is averaged over\textcolor{green}{{} }typically
1000 realizations of disorder $\{v_{n}\}$ and of the initial phases
$\arg[c_{n}(0)]$. Note that one has $p(g\rightarrow0,t=10^{5})<1$,
due to the fact that the initial state typically projects onto an
ensemble of localized eigenstates with localization lengths given
by eq.~(\ref{eq:LocalizationLenAnderson}). Some of these eigenstates
have localization lengths large enough to {}``touch'' the borders
of the box $\mathrm{L}$, so that the projection itself leads to losses.
One can see by comparing the three plots of fig.~\ref{fig:Lossesvsg}
that these losses decrease with the disorder $W$, and are independent
of $L_{0}$, as expected for the linear regime.

From fig.~\ref{fig:Lossesvsg} one can identify three main dynamical
regimes for increasing nonlinearity parameter $g$: i) A {}``quasi-localized''
dynamical regime in which localization is only weakly perturbed by
the nonlinearity; ii) a dynamical regime in which localization is
destroyed by the chaotic dynamics %
\footnote{The study of the site population dynamics, which displays in such
case continuous spectrum, confirms the presence of a chaotic behavior.%
}, the diffusion is reestablished and the losses increase notably;
iii) self-trapping regime in which diffusion (thus losses) is inhibited
again. It is interesting to note that, for concentrated initial states
$L_{0}\lesssim20$ and weak disorder, self-trapping can be much more
efficient than localization in suppressing diffusion. These dynamical
regimes correspond to those studied e.g. in Laptyeva \emph{et al.}.~\cite{Flach:DisorderNonlineChaos:EPL10}
by considering the long-term behavior of $\left\langle x^{2}\right\rangle $.
Fig.~\ref{fig:Lossesvsg} shows that the crossover between these
regimes strongly depends on the width of the initial state.

In the absence of nonlinearity, the initial state evolves until it
takes an exponentially localized shape of width $\sim\ell_{0}$ and
then {}``freezes''. In the presence of the nonlinearity, one observes
a global enlarging of the wavefunction, which results in a increasing
absorption at the borders of the box $\mathrm{L}$. In the weak disorder
limit we can construct an analytic model for these losses, starting
from the properties of the linear ($g=0$) system. This model, discussed
in appendix~\ref{app:Pnu}, produces an analytical expression describing
the asymptotic behavior of the survival probability $p(g=0,t)$, eq.~(\ref{eq:p(t)asympt}).
Remarkably, a small modification of this expression furnishes also
a function describing the asymptotic behavior \emph{in the nonlinear
case}, for values of $g$ as large as $10^{3}$. This expression is

\begin{widetext}\begin{equation}
p(g,t)=\frac{2}{\pi}\sin^{-1}\left(\sqrt{\frac{2\ell_{a}}{\ell_{\mathrm{eff}}\log(t/t_{2})+2\ell_{a}}}\right)\qquad(t\gg t_{2}),\label{eq:p(g,t)}\end{equation}
\end{widetext}where $\ell_{a}$ is given by eq.~(\ref{eq:la}),
depending on the geometry and on the localization length $\ell_{0}(W)$,
and $\ell_{\mathrm{eff}}$ and $t_{2}$ fitting parameters. With respect
to eq.~(\ref{eq:p(t)asympt}) of app.~\ref{app:Pnu} we simply replaced
$\ell_{0}$ by an {}``effective wavepacket length'' $\ell_{\mathrm{eff}}$
which is the second fitting parameter of our model. Fig.~\ref{fig:pvst_gneq0}
shows that the \emph{asymptotic} behavior of the survival probability
is very well fitted by this formula. The quantity $\ell_{\mathrm{eff}}$
plays a major role in the present work, as it can globally characterize
the wavepacket shape. We attribute this somewhat astonishing property
of our model to the fact that, even in presence of the nonlinearity,
the wavepacket displays exponential wings in the asymptotic regime,
as in can be verified numerically.

We display in fig.~\ref{fig:leffvsg} the behavior of the effective
wavepacket length $\ell_{\mathrm{eff}}$ as a function of the nonlinearity,
for disorder parameters $W=1$ {[}plot (a){]}, $W=2$ {[}plot (b){]}, and $W=3$ {[}plot (c){]}
and for various values of the initial state width $L_{0}$. One clearly
identifies the regions corresponding to the three dynamical regimes
discussed above: Quasi-localized regime, characterized by a constant
value of $\ell_{\mathrm{eff}}$, independent of $g$ and $L_{0}$;
chaotic regime, characterized by a marked increasing of $\ell_{\mathrm{eff}}$
with $g$ and strongly dependent on $L_{0}$; and self-trapping regime,
in which $\ell_{\mathrm{eff}}$ decreases again and can even become
smaller than the Anderson localization length $\ell_{0}$, and which
is also dependent on $L_{0}$. Fig.~\ref{fig:t2vsg} shows the behavior
of the {}``activation time'' $t_{2}$ (see App.~\ref{app:Pnu})
for the same set of states and parameters as fig.~\ref{fig:leffvsg}.
This time roughly corresponds the time for the wavepacket to {}``touch''
the border of the box $\mathrm{L}$, when losses become important.
In the quasilocalized regime $t_{2}$ is thus very high. The onset
of the chaotic behavior, favoring diffusion, produces a dramatic decrease
of $t_{2}$ which becomes virtually zero (that is, losses began almost
instantaneously). Self-trapping, inhibiting diffusion, produces an
increase of $t_{2}$. The same three dynamical regimes can hence be
observed also in the behavior of $t_{2}$. This fit parameter, although
necessary to obtain a good agreement with numerical simulation, thus
carries essentially the same information as $\ell_{\mathrm{eff}}$.
For $g\rightarrow\infty$, one expects the wave packet to evolve very
little due to {}``immediate'' self-trapping, that is, its asymptotic
width should be proportional $L_{0}$. In fig.~\ref{fig:leffvsg},
$\ell_{\mathrm{eff}}$ clearly has not attained this large $g$ regime,
but we observed numerically that even for moderate values of $g\sim300$
it is roughly proportional to $L_{0}$: $\ell_{\mathrm{eff}}(g=300)\approx0.25L_{0}$
for $W$ ranging from 2 to 4.

\section{Scaling laws\label{sec:ScalingLaws}}

The curves in fig.~\ref{fig:leffvsg} suggest the existence of scaling
laws. In fig.~\ref{fig:Scalinglaw} we show a plot of the quantity
$\widetilde{\ell}\equiv\ell_{\mathrm{eff}}L_{0}^{-3/4}$ as a function
of the scaled nonlinearity defined by\begin{eqnarray}
\widetilde{g} & \equiv & gL_{0}^{-3/4}\label{eq:g-tilda}\end{eqnarray}
using the same data of fig.~\ref{fig:leffvsg}. One observes, at
least for the $\widetilde{g}>1$, a clear grouping of the curves,
which indicates that the scaling applies essentially to the nonlinear
part of the behavior, as intuitively expected: Indeed, as $\ell_{\mathrm{eff}}(g=0)\rightarrow\ell_{0}$,
$\tilde{\ell}(g=0)\sim96W^{-2}L_{0}^{-3/4}$ which \emph{is dependent
of} $L_{0}$; this fact simply means that Anderson localization is
not controlled by $g$. On the nonlinearity-dominated region, however,
one\textcolor{green}{{} }would (naively) expect the nonlinear effects
to scale as $\left\langle v_{n}^{NL}\right\rangle \sim gL_{0}^{-1}$
for an initial state uniformly populating $L_{0}$ states. Presently,
we have no convincing explanation for the additional $L_{0}^{1/4}$
factor. We note that the scaling is not perfect for the small values
of $L_{0}$, for which the self-trapping is effective even for low
values of $g$.

Fig.~\ref{fig:Scalinglaw} shows that the scaling allows us to define
crossovers between the three dynamical regimes\emph{ }which\emph{
do not depend on} \emph{$L_{0}$}. This is confirmed in fig.~\ref{fig:leffvsgtildeW},
in which we plotted in false-colors the scaled effective wavepacket
length $\tilde{\ell}$ in the parameter plane $\tilde{g},W$ for $L_{0}=21$
{[}plot (a){]} and for $L_{0}=41$ {[}plot (b){]}. Despite of the
factor 2 in the width of the initial state, the two plots are almost
identical. The crossover between the quasi-localized regime and the
chaotic regime is found to be around $\tilde{g}_{c}\approx0.1$ and
the crossover between the chaotic regime and the self-trapping regime
around $\tilde{g}_{st}\approx5$. 

The use of these scaled variables thus allows us to characterize the
nonlinear dynamics independently of the size of the initial state,
which constitutes an important step in the understanding of these
complex dynamics.

\begin{figure}
\centering{}\includegraphics[width=0.7\columnwidth]{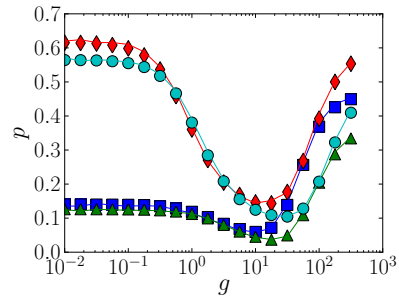}\caption{\label{fig:gaussian}Survival probability at $t=10^{5}$ for a square
initial state (full lines) and a gaussian initial state eq.~(\ref{eq:Gaussian})
(symbols), for $W=1,\: L_{0}=21$ (blue/squares), $W=1,\: L_{0}=41$
(green/triangles), $W=3,\: L_{0}=21$ (red/diamonds) and $W=3,\: L_{0}=41$
(cyan/circles). The dependence on the shape of the initial state is
very small.}
\end{figure}

\begin{figure}
\begin{centering}
\includegraphics[width=0.7\columnwidth]{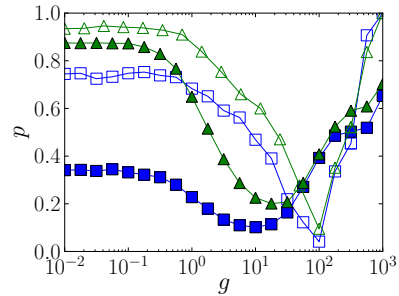}
\par\end{centering}

\caption{\label{fig:IncohvsCoh}Survival probability at $t=10^{5}$ for states
with coherent (empty symbols) and incoherent (full symbols) initial
phases, for $W=2,\, L_{0}=21$ (blue squares) and $W=4,\, L_{0}=21$,
(green triangles). The coherent case presents a marked enhancement
on the transport for $g\approx100$.}
\end{figure}

\begin{figure}
\begin{centering}
\includegraphics[width=0.7\columnwidth]{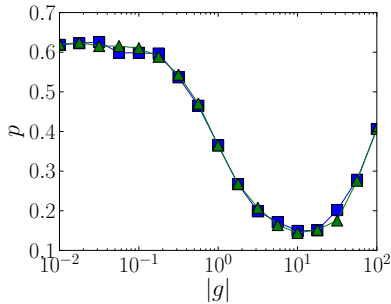}
\par\end{centering}

\caption{\label{fig:Sig_g}Effect of the sign of the nonlinearity with $L_{0}=21$
and $W=3$. The curves display the survival probability at $t=10^{5}$
for $g>0$ (blue squares) and $g<0$ (green triangles).}
\end{figure}

\section{Other effects\label{sec:OtherEffects}}

In this section, we consider three potentially significant effects
not taken into account above.

(i) \emph{The shape of the initial state}. In Fig.~\ref{fig:gaussian},
we compared the results discussed above, obtained with the {}``square''
initial state defined by eq.~(\ref{eq:InitialState}) with results
obtained with an initial gaussian state\begin{equation}
|c_{n}(0)|^{2}=\frac{1}{\sqrt{2\pi}\sigma}e^{-n^{2}/2\sigma^{2}}.\label{eq:Gaussian}\end{equation}
We set $\sigma=\sqrt{(L_{0}-1)(L_{0}+1)/12}$ so that, for a given
$L_{0}$, both square and gaussian initial states have the same second
momentum $\sum n^{2}|c_{n}|^{2}$. We see that the dynamics is independent
of the shape of the initial state to a very good approximation. In
the self-trapping regime, we see that the dynamics depends more on
the size $L_{0}$ than on the shape of the initial state.

(ii) \emph{The effect of the quantum phases of the initial state}.
Our choice of using initial random (incoherent) phases in the initial
state proved useful in allowing us to give a global characterization
of the dynamics. However, a particular coherent combination of $c_{n}$
phases can generate effects of quantum interference with an impact
on the dynamics. This is illustrated in Fig.~\ref{fig:IncohvsCoh},
where we compared the dynamics of an initial state of random phases
and an initial state with a constant phase for all $c_{n}$. The latter
one presents a different behavior, with a marked increasing in the
transport for $g\approx100$, which means that chaotic behavior is
highly favored in this case. Our previous result remains however a
valid description of the {}``average'' dynamics.

(iii) \emph{The effect of the sign of the nonlinearity}. In all the
results presented above $g>0$, that is, we have repulsive interactions.
In the limits of our approach, we have seen no significant difference
between attractive and repulsive interactions, see fig.~\ref{fig:Sig_g},
in contrast with some theoretical speculations. This confirms and
generalizes a result obtained in ref.~\cite{Rebuzzini:KRNonlinQuantumRes:PRE05}
in a different context.

Despite the simplifications we made in order to make this problem
tractable, it appears that our conclusions represent very well the
general behavior of the system, independently of most microscopic
parameters. We can thus say that we have characterized the dynamics
in a rather universal way.

\section{Conclusion}

Nonlinear dynamics is, in general, highly sensitive to initial conditions,
and no global characterization of the dynamics can be made unless
one can correctly take into account this dependence. This is particularly
true for the very important system studied in the present work: The
mean-field generalization of the Anderson model, described by the
Gross-Pitaevskii equation. We proposed a quantity providing a global
characterization of the dynamics independent of the size of the initial
state, the scaled effective wavepacket length $\tilde{\ell}$. This
is a first step in the necessary development of a coherent language
describing the effects of nonlinearities in Quantum Mechanics, which
will, most probably, constitute one of the major subjects of atomic
physics in the next years.
\begin{acknowledgments}
The authors are grateful to D. Delande and V. Zehnl\'e for fruitful
discussions. Laboratoire de Physique des Lasers, Atomes et Mol\'ecules
is UMR 8523 of CNRS. Work partially financed by the Agence Nationale
de la Recherche (MICPAF and LAKRIDI grants). 
\end{acknowledgments}
\appendix

\section{Determination of $p(g=0,t)$\label{app:Pnu}}

The aim of this appendix is to find an analytical formula describing
the evolution of the survival probability in the linear case $g=0$.
Let us first recall a few important results in the case of an infinite
lattice (no losses -- standard 1D Anderson model). In this case, the
corresponding Anderson Hamiltonian $H_{A}$ admits eigenstates $|\nu\rangle$
of energy $E_{\nu}$. In real space representation, $\psi_{\nu}(n)=\left\langle n\right.|\nu\rangle$
is centered at position $n_{\nu}$ in the lattice, with a localization
length $\ell_{\nu}\le\ell_{0}(W)$. As such states are exponentially
localized in the average, the density of presence of these eigenstates
at the site $n$ is

\begin{equation}
|\langle n|\nu\rangle|^{2}\approx\tanh\left(1/\ell_{\nu}\right)e^{-2|n-n_{\nu}|/\ell_{\nu}}.\label{eq:nnu}\end{equation}

In our problem we set an exponential imaginary potential (absorber)
on the border sites of the box $V_{i}(n)=-i\eta_{a}\exp\left(|n|/n_{c}\right)\quad\left(|n|>L/2\right)$.
In order to make the problem tractable, we make the assumption that
the imaginary potential can be characterize by a constant value $V_{i}=-i\eta_{a}$.
The corresponding Schr\"odinger equation is\[
i\frac{\partial|\psi(t)\rangle}{\partial t}=H_{A}|\psi(t)\rangle+V_{i}|\psi(t)\rangle.\]
Decomposing the wave function in the eigenfunctions basis, $|\psi(t)\rangle=\sum_{\nu}\alpha_{\nu}e^{-iE_{\nu}t}|\nu\rangle$,
we obtain (setting $\dot{\alpha}_{\nu}\equiv\partial\alpha_{\nu}/\partial t$):\[
i\dot{\alpha}_{\mu}=\sum_{\nu,n}\left\langle \mu\right|\left.n\right\rangle \left\langle n\right|\left.\nu\right\rangle V_{i}(n)\alpha_{\nu}\exp\left[i\left(E_{\mu}-E_{\nu}\right)t\right].\]
In the average, the contribution of $\mu\neq\nu$ is negligible and,
using eq.~(\ref{eq:nnu}), we get\[
i\dot{\alpha}_{\nu}=\left(\tanh\left(1/\ell_{\nu}\right)\sum_{n}V_{i}(n)\exp\left(-2\left|n-n_{\nu}\right|/\ell_{\nu}\right)\right)\alpha_{\nu}.\]

\begin{figure}
\centering{}\includegraphics[width=0.7\columnwidth]{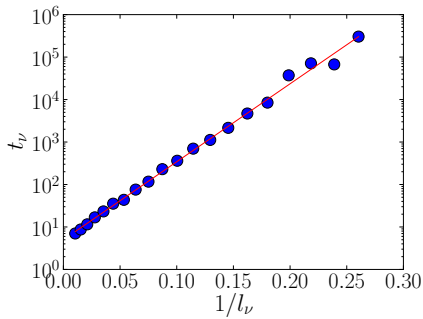}\caption{\label{fig:tnu}The blue circles\textcolor{green}{{} }correspond to
the values of $t_{\nu}$ obtained by fitting the numerically calculated
evolution of $\alpha_{\nu}$ with an exponential {[}eq.~(\ref{eq:alphanu}){]}.
The\textcolor{green}{{} }red solid line is a fit of these points by
eq.~(\ref{eq:tnu}).}
\end{figure}

Without loss of generality, we assume $n_{\nu}\ge\text{0}$ and note
$\ell_{a}$ the distance between the center of the wave packet corresponding
to this eigenstate and the closest border of the box at $n=L/2$.
Then \begin{eqnarray*}
\dot{\alpha}_{\nu} & = & -\eta_{a}\left(\tanh\left(1/\ell_{\nu}\right)\sum_{n\ge n_{\nu}+\ell_{a}}e^{-2(n-n_{\nu})/\ell_{\nu}}\right)\alpha_{\nu}\\
 & = & -\frac{\alpha_{\nu}\eta_{a}}{1+e^{-2/\ell_{\nu}}}e^{-2\ell_{a}/\ell_{\nu}}.\end{eqnarray*}
For moderate values of the disorder we can assume $1+e^{-2/\ell_{\nu}}\approx1$,
and obtain, integrating the above equation,\begin{eqnarray}
\alpha_{\nu} & = & \alpha_{\nu}(0)e^{-t/2t_{\nu}}\label{eq:alphanu}\\
t_{\nu} & = & \frac{1}{2\eta_{a}}e^{2\ell_{a}/\ell_{\nu}}.\nonumber \end{eqnarray}
We thus established two important results: (i) the attenuation (due
to the absorber) of an eigenstate is exponential, and (ii) the typical
time scale of absorption is proportional to $\exp\left(2\ell_{a}/\ell_{\nu}\right)$
\footnote{A similar form has been found in ref. \cite{Casati:QPoincareRes:PRL99}.%
}. These results can be generalized for an imaginary potential that
varies exponentially. We calculated numerically the evolution of $\alpha_{\nu}$
(with a non-constant imaginary potential) for a few eigenstates (averaged
over the disorder) and fitted this evolution with equation~(\ref{eq:alphanu})
to obtain the corresponding values of $t_{\nu}$, represented as circles
in Fig.~\ref{fig:tnu}, plotted as a function of $1/\ell_{\nu}$
in semi-log scale. The dependence of $t_{\nu}$ on $\ell_{\nu}$ can
in turn be fitted with the relation (red line in Fig.~\ref{fig:tnu})

\begin{eqnarray}
t_{\nu} & = & t_{2}\exp\left[2\ell_{a}\left(1/\ell_{\nu}-1/\ell_{0}(W)\right)\right].\label{eq:tnu}\end{eqnarray}
The fit parameter $t_{2}$ thus corresponds to the typical time-scale
for the attenuation of the eigenstate of largest width.

\begin{figure}
\begin{centering}
\includegraphics[width=0.7\columnwidth]{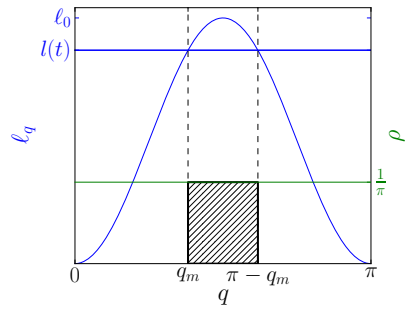}
\par\end{centering}

\caption{ \label{fig:schema}Graphical determination of $p$ in the weak-disorder
limit. The curve represents $\ell_{0}\sin^{2}q$ and its intersection
with $\ell(t)$ determines $q_{m}$. The dashed region correspond
to the integral eq.~(\ref{eq:p(t)int}).}
\end{figure}

\begin{figure}
\begin{centering}
\includegraphics[width=0.7\columnwidth]{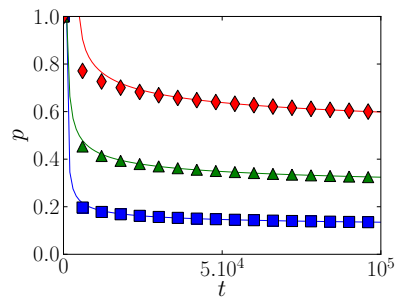}
\par\end{centering}

\caption{\label{fig:pvst_g=00003D0}Comparison between eq.~(\ref{eq:p(t)asympt})
and numerical simulations for $L_{0}=31$. The asymptotic behavior
of the survival probability is very well described by our formula,
for different values of the disorder, $W=1$ (blue squares), $W=2$
(green triangles), $W=3$ (red diamonds).}
\end{figure}

The evolution of the survival probability $p$ can be calculated for
a uniform initial state over $L_{0}$ Anderson states, i.e. $|\alpha_{\nu}(0)|^{2}=1/L_{0}$.
From eq.~(\ref{eq:alphanu})\[
p(g=0,t)=\sum_{\nu}|\alpha_{\nu}|^{2}=\frac{1}{L_{0}}\sum_{\nu}\exp(-t/t_{\nu})\]
with $t_{\nu}$ is given by eq.~(\ref{eq:tnu}). We now take $t_{\nu}$
as the typical time for decay of the eigenstate $\nu$, i.e. we approximate
$\exp(-t/t_{\nu})\approx\Theta(t_{\nu}-t)$, where $\Theta$ is the
Heaviside function: \[
p(g=0,t)=\frac{1}{L_{0}}\sum_{\nu}\Theta\left(t_{\nu}-t\right).\]
That is, we consider that the eigenstate is completely absorbed at
time $t_{\nu}$ at which it {}``touches'' the absorbing potential.
From eq.~(\ref{eq:tnu}) we see that

\[
\ell_{\nu}=\frac{2\ell_{a}}{\log\left(t_{\nu}/t_{2}\right)+2\ell_{a}/\ell_{0}}\]
At time $t$, the surviving states are thus those such that\[
\ell_{\nu}<=\ell(t)\equiv \frac{2\ell_{a}}{\log\left(t/t_{2}\right)+2\ell_{a}/\ell_{0}}.\]
In the weak-disorder limit one can use eq.~(\ref{eq:BandTB}) and
write \begin{equation}
p(g=0,t)=\int_{\ell_{q}<\ell(t)}\rho dq,\label{eq:p(t)int}\end{equation}
where $\rho=1/\pi$ is the density of states {[}eq.~(\ref{eq:DensityTB}){]}.
The condition $\ell_{q}<\ell(t)$ sets a maximum value $q_{m}$ of
the quasimomentum that can be determined\textcolor{green}{{} }graphically,
as shown in Fig.~(\ref{fig:schema}). The condition

\begin{eqnarray*}
\ell(t) & = & \ell_{0}\sin^{2}q_{m}\end{eqnarray*}
gives\[
q_{m}=\sin^{-1}\left(\ell(t)/\ell_{0}\right)^{1/2}\]
and, finally \begin{equation}
p(g=0,t)=\frac{2q_{m}}{\pi}=\frac{2}{\pi}\sin^{-1}\left(\frac{2\ell_{a}/\ell_{0}}{\log\left(t/t_{2}\right)+2\ell_{a}/\ell_{0}}\right)^{1/2}\label{eq:p(t)asympt}\end{equation}

Fitting the numerical survival probability with the above expression,
best results are obtained by setting

\begin{equation}
\ell_{a}=\frac{1}{2}\left(\frac{L}{2}-\frac{L_{0}}{4}-\frac{\ell_{0}}{5}\right),\label{eq:la}\end{equation}
as shown in Fig.~\ref{fig:pvst_g=00003D0}. One can think that the
distance $L/2-L_{0}/4$ is more likely to represent the typical distance
from the initial state to the absorber. The factor $1/2$ probably
compensates for the brutal step we made above of approximating the
exponential function by the Heaviside function. The presence of the
term in $\ell_{0}$ has a physical meaning: In our simulations, we
initially excite $L_{0}$ Wannier states, not $L_{0}$ Anderson states
and as a consequence, the initial size of the wave packet is larger
than $L_{0}$.


\end{document}